\pdfoutput=1
\documentclass[twocolumn, tighten]{aastex61}
\usepackage{amsmath}

\shorttitle{Optimized trajectories to the nearest stars}
\shortauthors{Heller, Hippke \& Kervella}

\begin{document}

\title{Optimized trajectories to the nearest stars using lightweight high-velocity photon sails}

\author{Ren\'{e} Heller}
\affiliation{Max Planck Institute for Solar System Research,
Justus-von-Liebig-Weg 3,
37077 G\"ottingen, Germany}
\email{heller@mps.mpg.de}

\author{Michael Hippke}
\affiliation{Luiter Stra{\ss}e 21b, 47506 Neukirchen-Vluyn, Germany}
\email{hippke@ifda.eu}

\author{Pierre Kervella}
\affiliation{Unidad Mixta Internacional Franco-Chilena de Astronom\'{i}a, CNRS/INSU UMI 3386 and Departamento de Astronom\'{i}a, Universidad de Chile, Casilla 36-D, Santiago, Chile}
\affiliation{LESIA (UMR 8109), Observatoire de Paris, PSL Research University, CNRS, Sorbonne Universit\'{e}s, UPMC Univ. Paris 06, Univ. Paris Diderot, Sorbonne Paris Cit\'{e}, 5 Place Jules Janssen, 92195 Meudon, France}
\email{pierre.kervella@obspm.fr}

\begin{abstract}
New means of interstellar travel are now being considered by various research teams, assuming lightweight spaceships to be accelerated {via either} laser or solar radiation to a significant fraction of the speed of light {($c$)}. We recently {showed} that gravitational assists can be combined with the stellar photon pressure to decelerate an incoming lightsail from Earth and fling it around a star or bring it to rest. Here, we demonstrate that photogravitational assist{s are} more effective when the star is used as a bumper (i.e. the sail passes ``in front of'' the star) rather than as a catapult (i.e. the sail passes ``behind''or ``around'' the star). This increases the maximum deceleration at $\alpha$\,Cen\,A and B and reduces the travel time of a nominal {graphene-class} sail {(}mass-to-surface ratio $8.6\times10^{-4}\,{\rm gram\,m}^{-2}$) from 95 to 75\,yr. The maximum possible velocity {reduction} upon arrival depends on the required deflection angle from $\alpha$\,Cen\,A to B and therefore on the {binary's} orbital phase. Here, we calculate the variation of the {minimum} travel times from Earth {into a bound orbit around} Proxima for the next 300\,yr {and then extend our calculations to roughly 22,000} stars within about 300\,ly. Although $\alpha$\,Cen is the most nearby star system, we find that Sirius\,A offers the shortest possible travel times {into a bound orbit}: 69\,yr assuming 12.5\%\,$c$ can be obtained at departure from the solar system. Sirius\,A thus offers the opportunity of flyby exploration plus deceleration into a bound orbit of the companion white dwarf after relatively short times of interstellar travel.
\end{abstract}

\keywords{radiation mechanisms: general --- solar neighborhood --- space vehicles --- stars: kinematics
and dynamics --- stars: individual ($\alpha$ Centauri, Sirius)}

\section{Introduction}

The possibility of interstellar travel has recently been revived through developments in laser technology, material sciences, and the construction of high-performance nano computer chips \citep{2016arXiv160401356L,2017ApJ...837L..20M}. The small weight of only a few grams of onboard equipment (communication, navigation, propulsion, science instruments, etc.) and the possible mass production of these probes results in reduced costs for production, launch, and operation -- benefits that could make such a mission affordable within the current century.\footnote{See the Breakthrough Starshot Initiative at\\ \url{http://breakthroughinitiatives.org}} Moreover, the first interplanetary solar sail mission \citep[IKAROS;][]{2011AcAau..69..833T} has been completed and further concepts are now being tested in near-Earth orbits \citep[e.g. LightSail;][]{Ridenoure2016}.

The idea of using solar light to accelerate a space probe in the solar system is not new \citep{1619dclt.book.....K,Tsander1924,Tsiolkovskiy1936,Garwin1958,Tsu1959}. {Key challenges are in the high temperatures close to the sun, where the thrust is strongest but the sail could melt \citep{Dachwald2005}, and} in the loss of effective propulsion at several AU from the Sun. Using a close (0.1\,AU) solar encounter and a 1000\,m radius sail, \citet{2008arXiv0809.3535M} calculated a maximum velocity of $0.2\%\,c$ for the departure of a 150\,kg probe from the solar system. Alternatively, it has been proposed that lasers could solve the decreasing-strength-with-distance problem due to their high flux of coherent light \citep{Forward1962,1966Natur.211...22M}. As noted by \citet{1967Natur.213..588R}, however, this launch technology meets the problem of deceleration at the destination since there would be no obvious way to decelerate the spacecraft at the target star.

\citet{2017ApJ...835L..32H} suggested to decelerate and deflect incoming high-velocity sails from Earth using the stellar radiation and gravitation, a maneuver they referred to as photogravitational assist. Assuming that the sail would be made of a strong, ultralight material such as graphene, which would be covered by a highly reflective broadband coating made of sub-wavelength thin metamaterials \citep{Slovick2013,Moitra2014}, such a sail could have a maximum speed at arrival ($v_{\infty, {\rm max}}$) of about $4.6\%\,c$ to be successively decelerated at the stellar triple $\alpha$\,Cen\,A, B, and C (Proxima) {\citep{2013JBIS...66..377M}}. Such a tour could potentially park the lightsail in a bound orbit around the Earth-mass habitable zone exoplanet Proxima\,b \citep{2016Natur.536..437A}. The travel times would be 95\,yr from Earth to $\alpha$\,Cen\,AB and another 46\,yr from the AB binary to Proxima, or 141\,yr in total.

Here, we present an alternative way of using photogravitational assists at $\alpha$\,Cen, which reduces travel times significantly. We also present a detailed analysis of the AB orbital motion for the next three centuries, which is crucial for a detailed study of the expected travel times at a given launch time from the solar system. We then apply this method to stars within 300\,ly of the sun and identify other highly interesting targets for bound-orbit exploration by interstellar lightsails.

\section{Methods}

\subsection{A Nominal Graphene-class Sail}

In our nominal scenario, we consider a sail made of a graphene structure ($\sigma=7.6{\times}10^{-4}\,{\rm gram\,m}^{-2}$) with a graphene-based rigid skeleton and highly reflective coating that is capable of transporting a science payload (laser communication, navigation, cameras, etc.) of about 1\,gram \citep{2017ApJ...835L..32H}. Such a sail must have an area of about $10^5\,{\rm m}^2~=~(316\,{\rm m})^2$ to make the weight of the science payload negligible against the weight of the sail structure. At this size, the graphene structure would contribute 76\,gram, the skeleton and coating could add 9\,gram, and the payload would add another 1\,gram, implying $\sigma_{\rm nom}=8.6{\times}10^{-4}\,{\rm gram\,m}^{-2}$ for our nominal graphene-class sail.

\subsection{Photogravitational Assists at $\alpha$\,Cen}

\subsubsection{An Improved Method for Deceleration}
\label{sub:improved}

In \citet{2017ApJ...835L..32H}, we showed how it is possible to use both the photon pressure of a star and its gravitational tug to decelerate and deflect an incoming lightsail. Our main aim was to determine the maximum possible injection speed ($v_{\infty, {\rm max}}$) at $\alpha$\,Cen\,A to allow a swing-by maneuver to $\alpha$\,Cen\,B and to finally achieve a bound orbit {around} Proxima. The key challenge that we identified for the determination of $v_{\infty, {\rm max}}$ is in reaching the maximum deceleration upon arrival at $\alpha$\,Cen\,A while simultaneously achieving the required deflection angle ($\delta$) between the inbound and outbound trajectories in order to swing from $\alpha$\,Cen\,A to B. Analytical estimates for our nominal graphene-class sail, which would approach the star as close as five stellar radii ($R_\star$), show that it could be possible to drop as much as $12,900\,{\rm km\,s}^{-1}$ at A, $8800\,{\rm km\,s}^{-1}$ at B, and another $1270\,{\rm km\,s}^{-1}$ at Proxima, giving an additive deceleration of up to $v_{\infty, {\rm max}}~=~22,970\,{\rm km\,s}^{-1}$ in total. 

We then performed numerical calculations using a modified $N$-body code that included the forces on the sail imposed by the stellar photon pressure. We imposed an analytic boundary condition on the sail's pitch angle ($\alpha$, the angle between the normal to the sail plane and the radius vector to the star) to maximize the loss of speed along its trajectory. In our simulations, the sail was inserted into the gravitational well of the star in such a way that it would pass the star on the one side \citep[e.g. on the right, see Fig.~2 in][]{2017ApJ...835L..32H} and have a net deflection after its passage to the other side (e.g. to the left).

These numerical simulations revealed that the maximum velocity upon arrival at A, which would allow a deflection to B (assuming $\delta~{\approx}~10^\circ$), is limited to $v_{\infty, {\rm max}}=13,800\,{\rm km\,s}^{-1}$. And so while analytical estimates of the additive nature of the photogravitational assist suggest that up to $v_{\infty, {\rm max}}~=~22,970\,{\rm km\,s}^{-1}$ could be successively absorbed at the $\alpha$\,Cen stellar triple, numerical simulations show that the particular geometry of the system limits $v_{\infty, {\rm max}}$ to $13,800\,{\rm km\,s}^{-1}$.

\begin{figure}
\includegraphics[width=\linewidth]{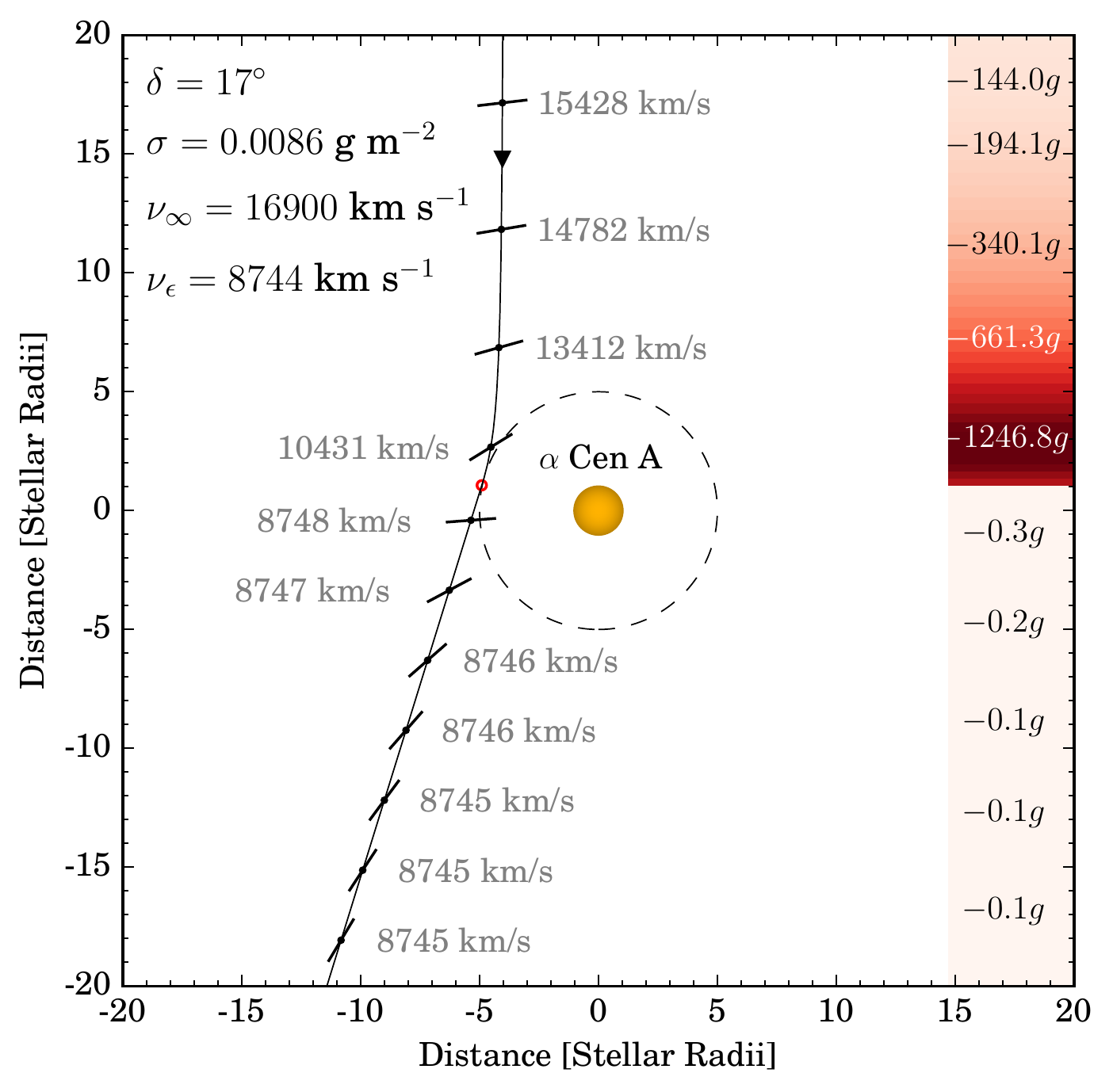}
\caption{\label{fig:pga} Trajectory of a lightsail performing a photogravitational assist at $\alpha$\,Cen\,A (orange circle). The bars along the trajectory visualize the instantaneous orientation of the sail {(in steps of 60\,min)} determined to maximize the deceleration. The values in the legend denote the deflection angle, the mass-to-surface ratio, the inbound velocity, and the outbound velocity of the sail. The color bar at the right shows the $g$-forces acting on the sail along the trajectory, where $g~=~9.81\,{\rm m\,s}^{-2}$ is the acceleration on the Earth's surface.}
\end{figure}

\begin{figure}
\centering
\includegraphics[width=.68\linewidth]{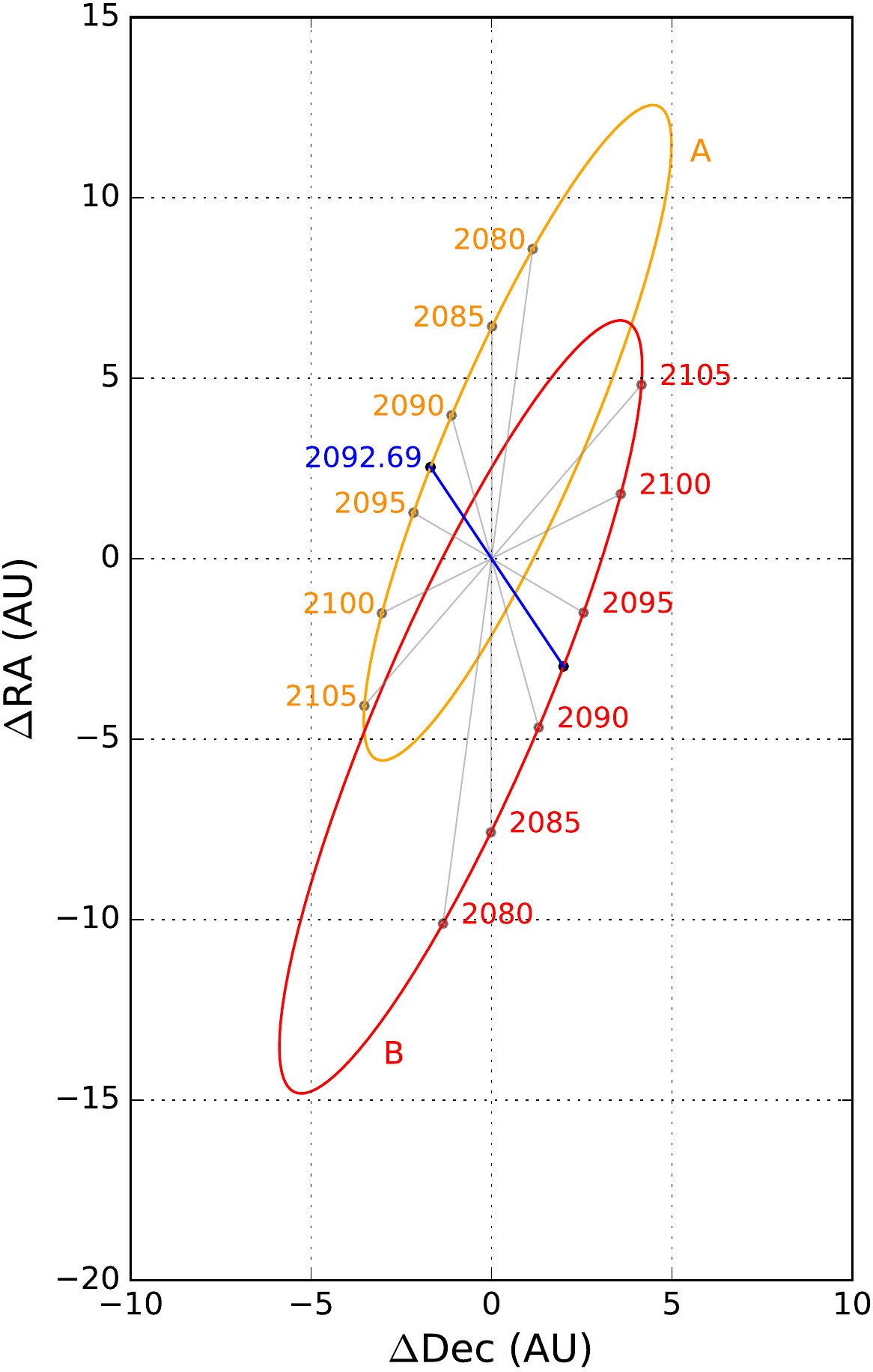}
\caption{\label{fig:alphaCensky} Orbital trajectories of $\alpha$\,Cen\,A (orange) and B (red) in their barycentric coordinate system as seen from Earth using differential R.A. and decl. coordinates. The AB vector at the time of their closest apparent encounter in 2092.69 (8 September 2092) is marked in blue.}
\end{figure}

Following up on these simulations, we recently discovered that a simple modification of the incoming trajectory yields higher deflection angles at even higher incoming speeds. It turns out that it is more effective to use the stellar photon pressure (rather than gravity) to enhance the deflection. From a geometry perspective, it is more efficient to let the sail approach the star on the same side (e.g. on the left; see Figure~\ref{fig:pga}) as the desired deflection (i.e. to the left). We then find that, using the same optimization strategy for deceleration as \citet{2017ApJ...835L..32H}, a maximum total deceleration of $v_{\infty, {\rm max}}~=~17,050\,{\rm km\,s}^{-1}$ can be reached at $\delta=19^\circ$, where $8800\,{\rm km\,s}^{-1}$ and $8400\,{\rm km\,s}^{-1}$ can be lost at A and B, respectively. If the lightsail is supposed to continue its journey on to Proxima, then it would actually be better to orient the sail during its passage at B in a way to avoid maximum deceleration, so that the sail can continue its cruise to Proxima with a residual speed of $1270\,{\rm km\,s}^{-1}$. This is the maximum speed that can ultimately be absorbed at Proxima.

\subsubsection{A Timetable of Launch Opportunities to Proxima}

The orbital motion of the $\alpha$\,Cen\,AB binary leads to a periodic variation \citep[$P=79.929\pm0.013$\,yr;][]{2016A&A...594A.107K} of the deflection at $\alpha$\,Cen\,A that is required by the sail to reach B. We calculate the {binary's} orbital motion for the next 300\,yr, based on the works of \citet{2016A&A...594A.107K}, and we include the computation of Proxima's orbit around the $\alpha$\,Cen\,AB binary based on \citet{2017A&A...598L...7K}. Figure~\ref{fig:alphaCensky} shows the orbits of $\alpha$\,Cen\,A (orange) and B (red) as a projection on the sky with dates around the year 2100 labeled along the ellipses. The blue line illustrates the sky-projected AB vector at the time when the deflection required by the sail for a sequential AB photogravitational assist is smallest. This event will take place on 8 September 2092 (2092.69).

We then use our results for $\delta(t)$, where $t$ symbolizes time, together with a range of numerical trajectory simulations, to first determine $v_{\infty, {\rm max}}(\delta)$, then $v_{\infty, {\rm max}}(t)$, that is, the maximum possible injection speed at $\alpha$\,Cen to reach a bound orbit at Proxima, and ultimately the travel time from Earth to $\alpha$\,Cen\,AB over the next 300\,yr. For the final step, the conversion from travel speed to travel time, we adopt a barycentric distance to $\alpha$\,Cen\,AB of $4.365$\,ly \citep{2016A&A...594A.107K}.

\subsection{An Interstellar Travel Catalog}

\subsubsection{Analytical Estimates}

\begin{figure}
\centering
\includegraphics[width=1\linewidth]{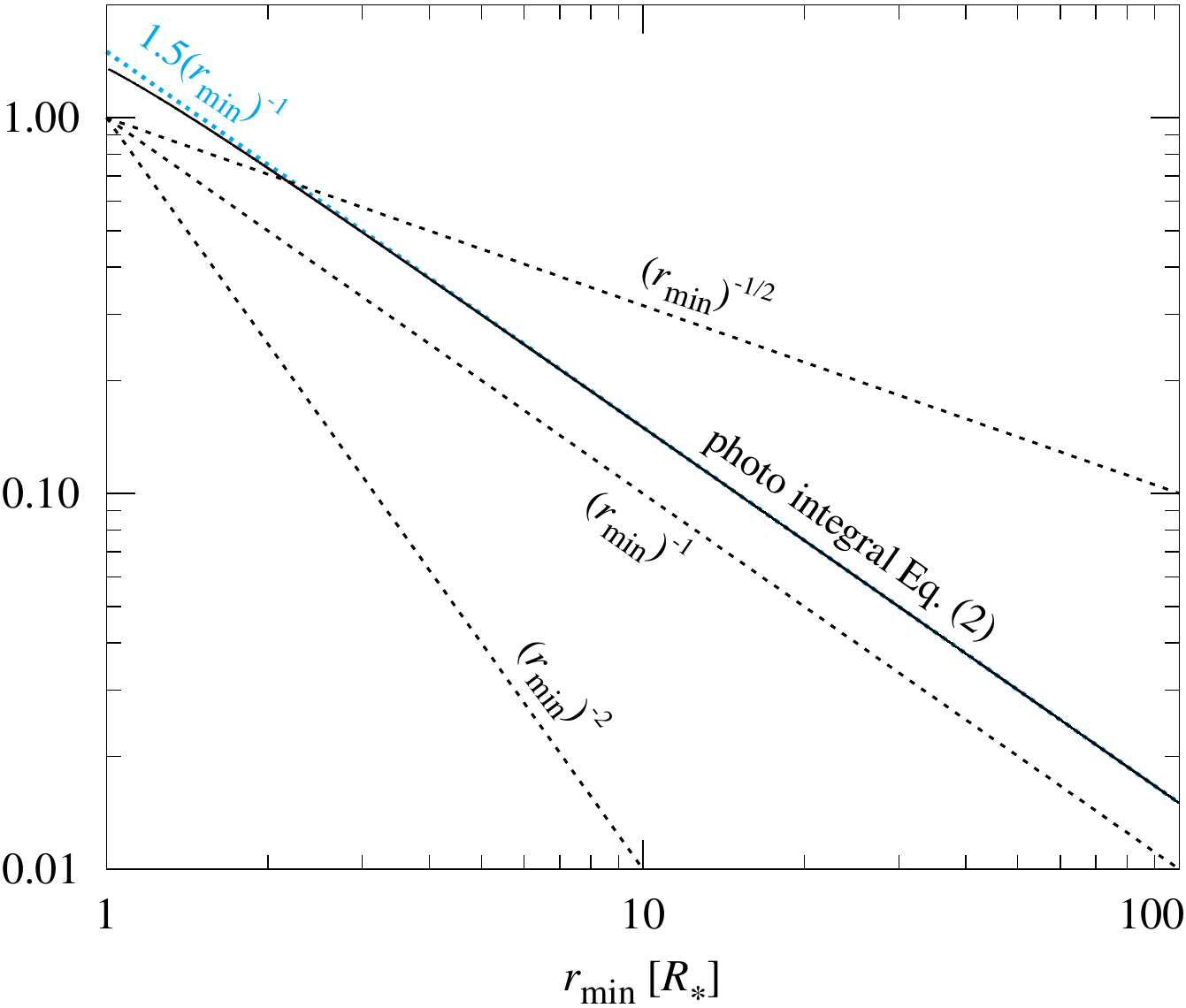}
\caption{\label{fig:photointegral} {Integral in Equation~\eqref{eq:Ekin1} as a function of $r_{\rm min}$, the closest stellar encounter (solid line). Our nominal calculations invoke a sail approaching as close as $r_{\rm min}=5\,R_\star$, where the integral amounts to 0.299. Four different functions are shown for comparison (dotted lines). Note that $1.5/r_{\rm min}$ provides an exquisite fit (blue dotted line), e.g. 1.5/5~=~0.3.}}
\end{figure}

While the $\alpha$\,Cen system is the natural first target for interstellar travels to consider, given its proximity and the presence of the Earth-sized, potentially habitable planet Proxima\,b, other nearby stars offer compelling opportunities, too. We thus extend our investigations of photogravitational assists to a full stop around other nearby star{s and} estimate the travel time ($\tau$) to a nearby star with a given radius ($R_\star$) and luminosity ($L_\star$) as

\vspace{-.0cm}

\begin{equation}\label{eq:tau}
\tau = \frac{d}{v_{\infty,{\rm max}}} = \frac{d}{\displaystyle \sqrt{\frac{2E_{\rm kin}}{M}}} \ ,
\end{equation}

\noindent
where $d$ is the stellar distance to the Sun, $M$ is the sail mass,

\vspace{-.50cm}

\begin{align}\label{eq:Ekin1}
E_{\rm kin,p} &=  {\displaystyle \int\limits_{r_{\rm min}}^{\infty} } dr F(r) \\ \nonumber
                      &= \frac{\displaystyle L_\star A}{\displaystyle 3{\pi}cR_\star} \ {\displaystyle \int\limits_{r_{\rm min}}^{\infty}} dr {\Bigg (} {\Bigg [} 1- {\Big [}1- {\Big (} \frac{\displaystyle R_\star}{\displaystyle r} {\Big )}^2{\Big ]}^{3/2} {\Bigg ]} \frac{1}{R_\star} {\Bigg )}
\end{align}

\noindent
is the kinetic energy of the sail that can be absorbed by the stellar photons {during approach} \citep{1990CeMDA..49..249M,2017ApJ...835L..32H}, and $A$ is the sail's surface area. The integral in Equation~\eqref{eq:Ekin1}, which we refer to as the photointegral, is independent of the actual value of $R_\star$ {and only depends on the choice of $r_{\rm min}$. Figure~\ref{fig:photointegral} shows the numerical value of the photointegral for $1\,R_\star\,{\leq}\,r_{\rm min}\,{\leq}\,100\,R_\star$. Four different functions are given for comparison (dotted lines). In particular, we found that $1.5/r_{\rm min}$ (with $r_{\rm min}$ in units of $R_\star$) provides an excellent approximation with deviations $<~1\%$ for $r_{\rm min}~\geq~3\,R_\star$ (blue dotted line). Using this approximation, Equation~\eqref{eq:Ekin1} becomes}

\vspace{-.20cm}

\begin{align}\label{eq:EkinGeneral}
E_{\rm kin,p} &= \frac{\displaystyle L_\star A}{\displaystyle 2{\pi}cR_\star (r_{\rm min}/R_\star)} = \frac{\displaystyle L_\star A}{\displaystyle 2{\pi}c \, r_{\rm min}}
\end{align}

\vspace{.10cm}

\noindent
{and Equation~\eqref{eq:tau} then is equivalent to}

\vspace{-.20cm}

\begin{align}\label{eq:logspirGeneral}\nonumber
                       & \hspace{1.9cm}   \tau = \frac{d}{ \sqrt{   \frac{\displaystyle L_\star A}{\displaystyle {\pi}c \, r_{\rm min}M} }  } \\
\Leftrightarrow & \log_{10}  {\Big (}  \tau({\rm yr})  {\Big )} = \frac{1}{\ln(10)} \ln {\Bigg (}  \frac{d/{\rm yr}}{      \sqrt{   \frac{\displaystyle L_\star A}{\displaystyle {\pi}c \, r_{\rm min}M}  }   }  {\Bigg )}   \hspace{.5cm} .
\end{align}

We chose the latter equivalent transformation to describe the travel time as a logarithmic spiral of the form $\varphi(d)=1/k \ln(d/a)$, where the constant $k>0$ defines the curvature of the spiral and $a$ is the radius of the circle for $k~{\rightarrow}~0$. As an aside, note that the first line in Equation~\eqref{eq:logspir} is equivalent to $\tau~{\propto}~\sigma$, a relation that we will come back to below.

For a nominal minimum stellar approach of $r_{\rm min}~=~5\,R_\star$, the photointegral yields a value of $0.299$ {and Equation~\eqref{eq:Ekin1} collapses to}

\vspace{-.20cm}

\begin{align}\label{eq:Ekin2}
E_{\rm kin,p} &= \frac{\displaystyle L_\star A}{\displaystyle 10{\pi}cR_\star} \hspace{.5cm} , \hspace{.5cm} {\rm for} \ r_{\rm min}=5\,R_\star \ .
\end{align}

{so that}

\begin{equation}\label{eq:logspir}
\log_{10}  {\Big (}  \tau({\rm yr})  {\Big )} = \frac{1}{\ln(10)} \ln {\Bigg (}  \frac{d/{\rm yr}}{      \sqrt{   \frac{\displaystyle L_\star A}{\displaystyle 5{\pi}cR_{\star}M}  }   }  {\Bigg )}
\end{equation}

\subsubsection{Numerical Simulations}

In addition to our analytical estimates, we perform numerical simulations of photogravitational assists around stars in the solar neighborhood as in \citet{2017ApJ...835L..32H}. For 117 stars within 21\,ly around the Sun, we use distances, luminosities, masses and radii from \citet{2004A&A...420..183A}, \citet{2005ApJS..159..141V}, \citet{2007A&A...474..653V}, and \citet{2009A&A...501..941H}.\footnote{For an extensive list of references, see \\ \url{http://www.johnstonsarchive.net/astro/nearstar.html}} 

For more distant stars but within 316\,ly, we use parallax measurements from Hipparcos \citep{1997A&A...323L..49P} and Gaia DR1 \citep{2016A&A...595A...2G,2016A&A...595A...1G,2016A&A...595A...4L} to estimate stellar distances. We pull available estimated temperatures from the Radial Velocity Experiment (RAVE) \citep{2017ApJ...840...59C,2017AJ....153...75K} and add additional stars from the Hipparcos and Tycho-2 catalog \citep{2000A&A...355L..27H}, where only color information ($B-V$) is available. For these latter stars, we follow the procedure of \citet{2009A&A...508.1509H} to estimate the effective temperature as

\begin{equation}\label{equ:Teff_BV}
T_\mathrm{eff} = 10^{[14.551 - (B-V)] / 3.684}\,\mathrm{K} \hspace{0.2cm} ,
\end{equation}

\noindent
the stellar radius $R_\star$ as

\begin{equation}
  \frac{R_\star}{R_\odot} = \left[ \left( \frac{5\,770\,\mathrm{K}}{T_\mathrm{eff}} \right)^4 10^{(4.83 - M_\mathrm{V})/2.5} \right]^{1/2} \hspace{0.2cm} ,
\end{equation}

\noindent
and the stellar mass $M_\star$ as

\begin{equation}\label{equ:mlr}
  \frac{M_\star}{M_\odot} = \left( \frac{4 \pi R_\star^2 \sigma_{\mathrm{SB}} T_\mathrm{eff}^4}{L_\odot}\right)^{1 / \beta} \hspace{0.2cm} ,
\end{equation}

\noindent
where $\sigma_{\mathrm{SB}}$ is the Stefan-Boltzmann constant and where the coefficient $\beta$ in the relation $L~\propto~M^\beta$ depends on the stellar mass \citep[see Table~\ref{tab:beta}; values taken from][]{1983Ap&SS..96..125C}.

\startlongtable
\begin{deluxetable}{ccc}
\tablecaption{\footnotesize Empirical Values for $\beta$ in the Mass-Luminosity Relation. \label{tab:beta}}
\tablehead{ \colhead{ $\beta$ } & & \colhead{Stellar Mass} } 
\startdata
    $3.05 \pm 0.14$ & \hspace{1.7cm} \ \ \ & $M_\star \lesssim 0.5\,M_\odot$  \\
    $4.76 \pm 0.01$ &  & $0.6\,M_\odot \lesssim M_\star \lesssim 1.5\,M_\odot$  \\
    $3.68 \pm 0.05$ &  & $1.5\,M_\odot \lesssim M_\star$  \\
\enddata
\end{deluxetable}

In our modified $N$-body simulations, we impose an upper temperature ($T$) limit of $100^{\circ}$C (373\,K) on the sail. At these temperatures, modern silicon semiconductors are still operational \citep{Intel} \, and the sail material is likely not a limitation. For comparison, aluminum has a melting point of 933\,K, and graphene melts at 4510\,K \citep{2015PhRvB..91d5415L}.

We assume a sail reflectivity of 99.99\%, which might be achievable in the broadband using multiple coatings, or metamaterials with sub-wavelength thickness \citep{2013PhRvB..88p5116S, 2014ApPhL.104q1102M, 2016JPhD...49s5101L}. From the Boltzmann law, we first deduce

\begin{equation}\label{eq:heating}
T_{\rm eff} < T \times (n^2/\zeta)^{1/4} \ ,
\end{equation}

\noindent
where $\zeta= (1-0.9999)$ is the absorptivity of the sail and $n~=~r_{\rm min}/R_\star$, and then we calculate

\begin{equation}
n = \sqrt{\zeta}\times(T_{\rm eff}/T)^2
\end{equation}

\noindent
as the minimum (float) number of stellar radii for the sail to prevent heating above $T=373$\,K. As an example, for Sirius\,A ($T_{\rm eff}~=~8860$\,K) we obtain $n~=~\sqrt{1-0.9999}\times(8860/373)^2 = 5.6$ (stellar radii) as a minimum distance. For stars with $T_{\rm eff}~<~8340$\,K we have $n~<~5$ and so we impose $n~=~5$ to limit the destructive perils of flares, magnetic fields, electron/proton impacts, etc.

\section{Results}

\begin{figure*}[ht]
\includegraphics[width=\linewidth]{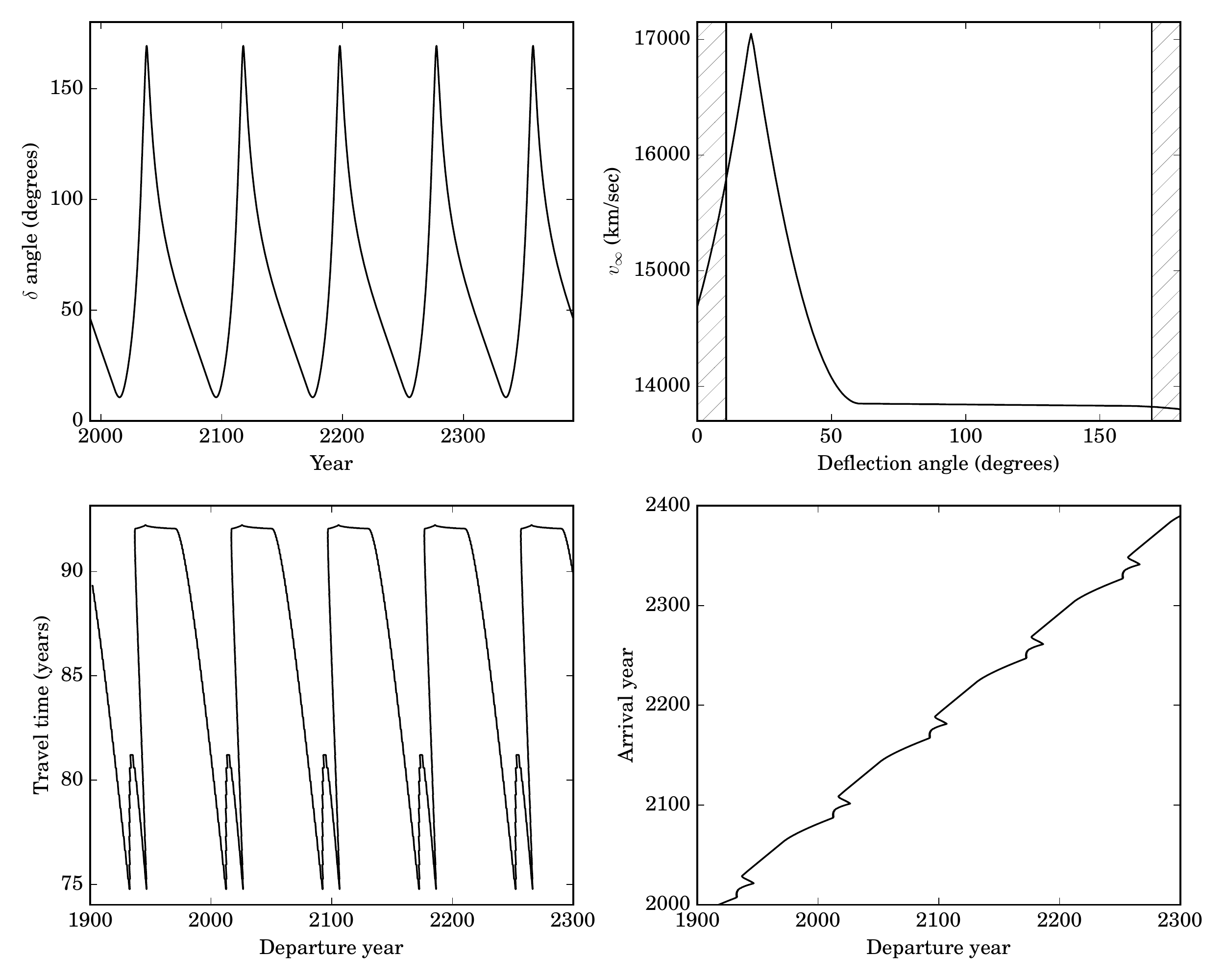}
\caption{\label{fig:alphaCen} Deflection angles between $\alpha$\,Cen\,A and B and maximum injection speeds at $\alpha$\,Cen required by a nominal lightsail made of graphene to perform photogravitational assists into a bound orbit around Proxima. Values have been determined by numerical trajectory simulations with minimum stellar approaches of $5\,R_\star$ around either $\alpha$\,Cen\,A or B, whichever was closer to Earth at the time of arrival. {\it Top left}: Temporal variation of the angular deflection required upon arrival at $\alpha$Cen to either reach B after the encounter with A or to reach A after encounter with B.  {\it Top right}: Maximum possible injection speed at $\alpha$\,Cen\,A to achieve a given deflection angle. The maximum speed can be obtained for $\delta=19^\circ$. {\it Bottom left}: Year of departure from Earth versus travel time between Earth and $\alpha$\,Cen\,AB. The pairs of depressions correspond to maximum injection speeds of $17,050\,{\rm km\,s}^{-1}$ at angular separations of $19^\circ$ (see upper right panel), which occur a few years before and after the closest encounter of the $\alpha$\,Cen\,AB binary. {\it Bottom right}: Arrival time at $\alpha$\,Cen\,AB as a function of departure time from Earth.}
\end{figure*}

\begin{figure*}
\includegraphics[width=1.\linewidth]{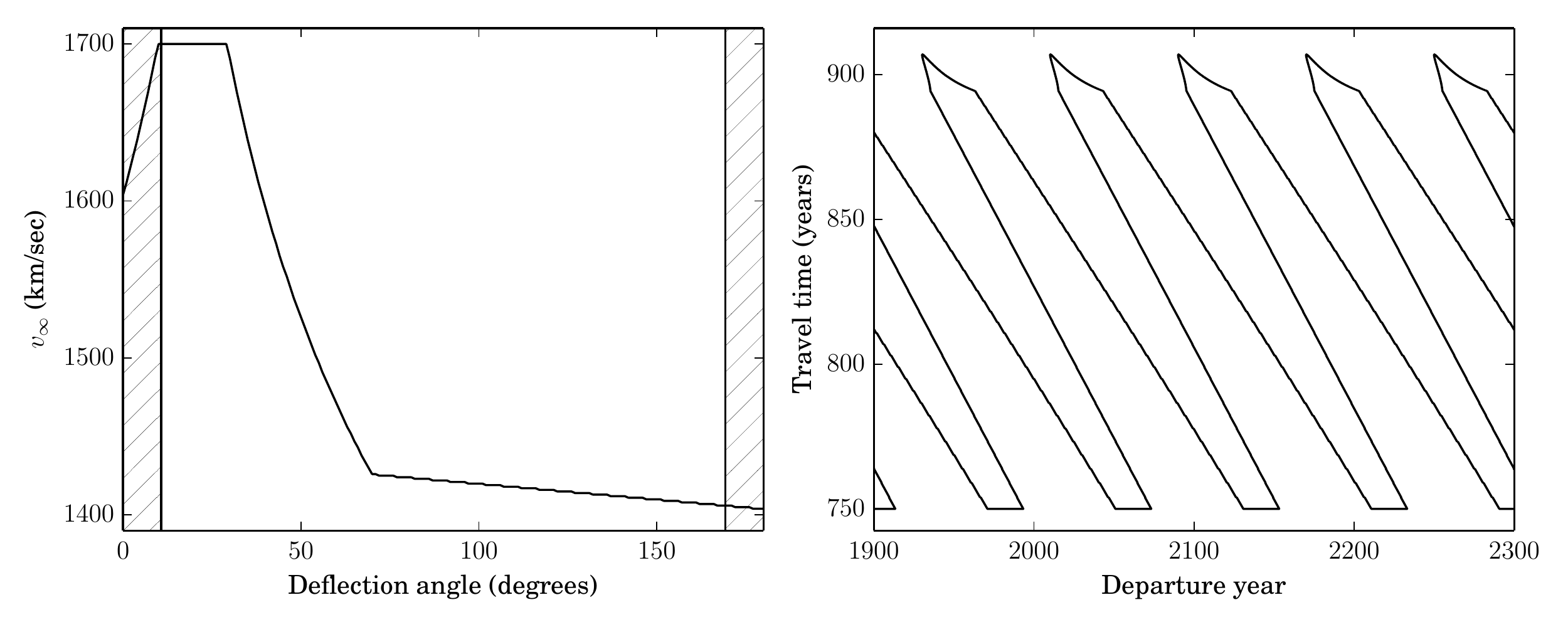}
\caption{\label{fig:alu} Same as top right and bottom left panels in Figure~\ref{fig:alphaCen}, but now for an aluminum lattice sail. The maximum injection velocities (left panel) are about a factor of 10 smaller for a given angle than in for the case of a graphene-class lightsail. Consequently, the corresponding travel times are a factor of 10 longer (right panel); see Equation~\eqref{eq:logspir}.}
\end{figure*}

\subsection{Optimized Trajectories to Proxima}
\label{sub:optimized}

Using photogravitational assists at the same side of the star as the desired deflection (see Figure~\ref{fig:pga}) rather than gravitational swing-bys ``behind'' the star \citep[as in][]{2017ApJ...835L..32H}, we find that the maximum possible injection speed at $\alpha$\,Cen\,A to allow deflections to B and then to C can be significantly increased. For a nominal graphene-class lightsail with a mass-to-surface ratio of $\sigma_{\rm nom}=10~{\times}~8.6\,{\rm gram\,m^{-2}}$, we find $v_{\infty, {\rm max}}=17,050\,{\rm km\,s}^{-1}$ ($5.7\%\,c$). Compared to the previously published value of $v_{\infty, {\rm max}}=13,800\,{\rm km\,s}^{-1}$ \citep[$4.6\%\,c$, ][]{2017ApJ...835L..32H}, this corresponds to an increase of 24\% in speed and implies a reduction of the travel time from Earth to $\alpha$\,Cen\,AB from 95\,yr to 75\,yr. The total travel time from Earth to a full stop at Proxima then becomes $75\,{\rm yr}~+~46\,{\rm yr}~=~121$\,yr, assuming a residual velocity of $1280\,{\rm km\,s}^{-1}$ can be absorbed at Proxima after 46\,yr of travel between $\alpha$\,Cen\,B and Proxima.

In Figure~\ref{fig:alphaCen}, we present our results for the variation of the total travel time to Proxima over the next 300\,yr. The upper left panel shows the deflection angle required by the sail upon passage of $\alpha$\,Cen\,A to reach B. The minimum value is $10.7^\circ$.

In the top right panel, we show the maximum possible injection speed at A that allows deflection by an angle $\delta$. The peak velocity of $17,050\,{\rm km\,s}^{-1}$ occurs at an angle of $19^\circ$. The shaded regions in the panel denote angles smaller (or larger) than the smallest (or largest) angular separation between A and B, which are thus irrelevant for real trajectories. 

The bottom left panel shows the variation of the total travel time from Earth to $\alpha$\,Cen\,A given our knowledge about the upcoming orbital alignments (top left panel) and the possible maximum injection speeds (top right panel). The pairs of depressions, e.g. near the departure years 2093 and 2106, correspond to phases of close alignments between A and B, with the first depression referring to an A--B sequence and the second depression corresponding to a B--A sequence, where B is visited first and A thereafter (since A in this case is behind B, as seen from Earth).

\begin{figure*}
\centering
\includegraphics[width=.475\linewidth]{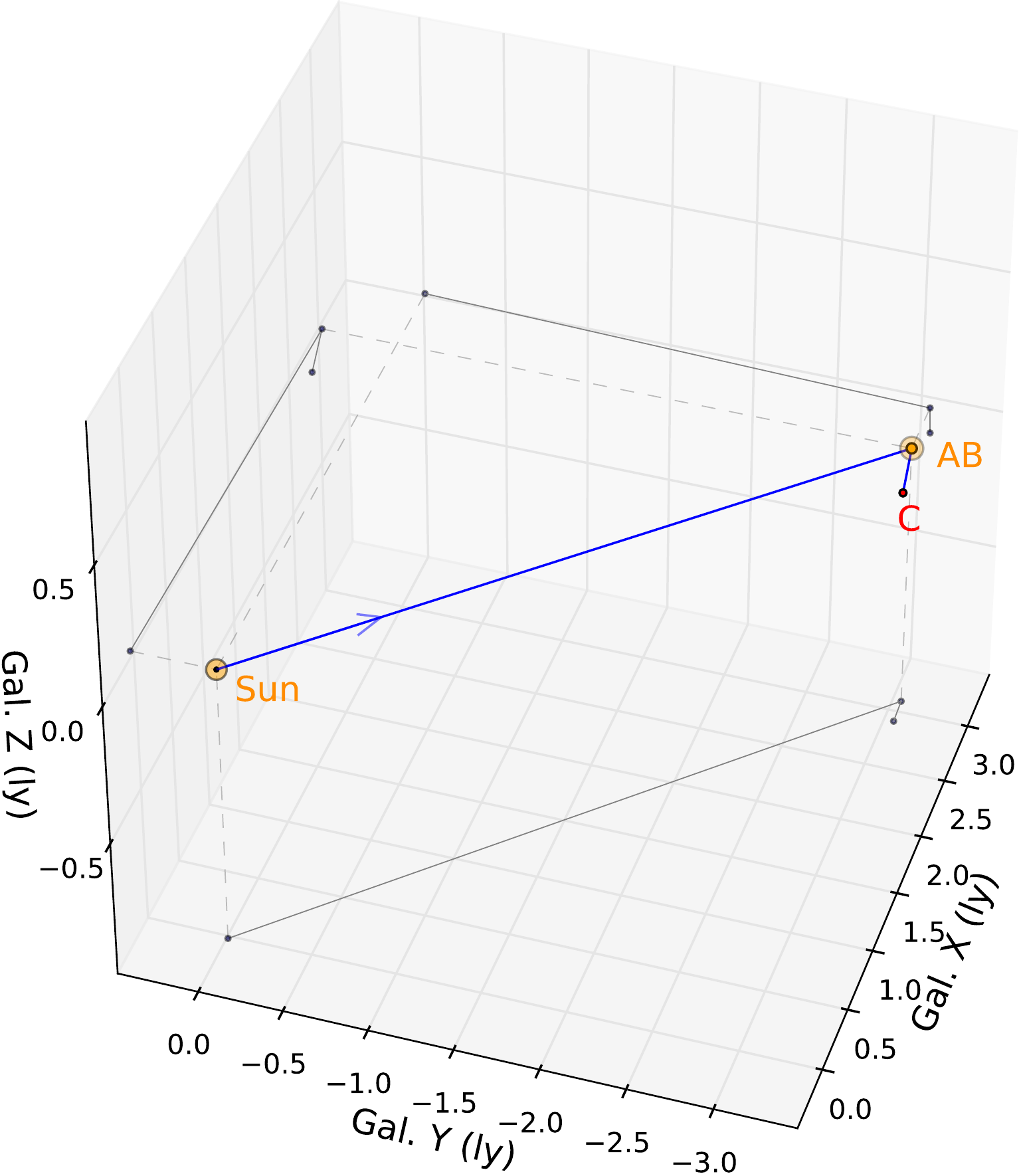}
\hspace{.32cm}
\includegraphics[width=.49\linewidth]{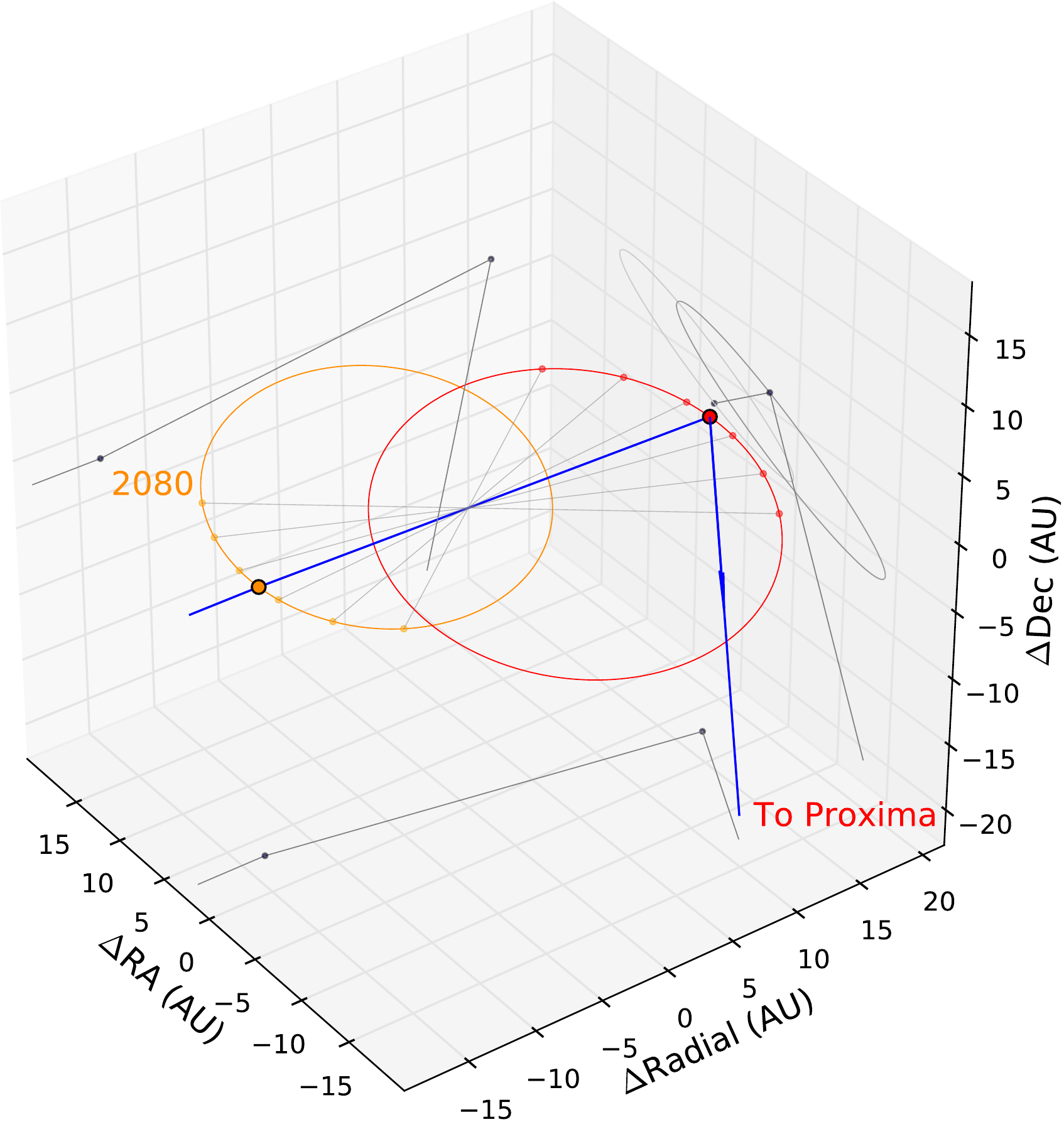}
\caption{\label{fig:trajectory} Example trajectory (blue line) of a lightsail from Earth performing photogravitational assists at $\alpha$\,Cen\,A and B toward Proxima. Projections of the trajectory on the three planes of the coordinate systems are shown as gray lines. {\it Left}: Large-scale overview of the trajectory from the Sun to $\alpha$\,Cen in Galactic coordinates (in units of ly). $X$ increases toward the Galactic center, $Y$ is positive toward the Galactic direction of rotation, and is $Z$ positive toward the north Galactic pole. {\it Right}: Orbital configuration of the $\alpha$\,Cen\,AB binary upon arrival of the lightsail in 2092.69. The origin of the differential cartesian ICRS coordinate system (in units of AU) is located in the $\alpha$\,Cen\,AB barycenter. The points on the orbits of A (orange ellipse) and B (red ellipse) are separated by 5\,yr to illustrate the evolution of the stellar positions. The projection of the binary orbit on the Earth sky is shown in the R.A.-decl. plane.}
\end{figure*}

The bottom right panel, finally, shows the year of arrival given a year or departure and assuming maximum injection speeds at the time of arrival. As a bottom line of this analysis, we find that the minimum travel time of a graphene-class lightsail from Earth to the $\alpha$\,Cen\,AB binary varies between about 75\,yr and 95\,yr. These values will reduce by a factor of about 3/4 over the next roughly $30,000$\,yr as the $\alpha$\,Cen system approaches the Sun from a bit more than 4\,ly today to a minimum separation of about 3\,ly.

In Figure~\ref{fig:alu}, we extend our study and investigate other possible mass-to-surface ratios for the lightsail. The two panels show the maximum injection speed as a function of the deflection angle (left) and the travel time from Earth to $\alpha$\,Cen\,AB as a function of time for an aluminum lattice sail as proposed by \citet{Drexler1979}. The corresponding mass-to-surface ratio is $\sigma~=~7~\times~10^{-2}\,{\rm gram\,m}^{-2}$. The results from these numerical simulations suggest that $v_{\infty, {\rm max}}$ is about a factor of 10 smaller for any given deflection angle than for our nominal graphene-class sail and that the travel time is about a factor of 10 higher. This is consistent with our analytical estimate that $\tau~{\propto}~\sigma$ (see Equation~\ref{eq:logspirGeneral}), which suggests travel times of an aluminum lattice are extended by a factor of $\sqrt{7\times10^{-2} / 7.6\times10^{-4}} = 9.7$ compared to a graphene-class sail.

The right panel of Figure~\ref{fig:alu} covers times of departure from Earth between 1900 and 2300 as the bottom right panel of Figure~\ref{fig:alphaCen}. Minimum travel times for an aluminum-class lightsail are $750\,{\rm yr}~{\lesssim}~\tau(t)~{\lesssim}~910$\,yr. Note that any given departure year from Earth allows multiple, in fact, up to five possible travel times times. Each individual travel time corresponds to a specific, sub-$v_{\infty, {\rm max}}$ interstellar speed of the lightsail and a specific orbital cycle of the AB binary upon arrival.

Figure~\ref{fig:trajectory} illustrates the complete trajectory of a lightsail from Earth to $\alpha$\,Cen\,A, B, and C with an arrival at $\alpha$\,Cen\,AB in the year 2092.69. At that instance, the sky-projected AB separation is $19^\circ$ and $v_{\infty, {\rm max}}~=~17,050\,{\rm km\,s}^{-1}$ is possible, providing the minimum travel time from Earth of 75\,yr (see Figure~\ref{fig:alphaCen}, bottom left). The left panel shows a global overview of the Sun--AB--C trajectory in Galactic coordinates and on a scale of light years. Our nominal graphene-class sail would require a minimum of 75\,yr to cover the distance of the long blue line between the Sun and the AB binary and another 46\,yr to complete the travel from the AB binary to C. The right panel shows a zoom into the AB binary at the time of arrival in 2092.69. With an instantaneous separation of 30.84\,AU between A and B and assuming a residual speed of the sail of $8400\,{\rm km\,s}^{-1}$ (see Section~\ref{sub:improved}), the travel time between encounters of A and B would be 6 days and 8.6 hours. After the encounter with B, the lightsail would be deflected toward Proxima. A projection of the blue trajectory on the Earth sky is shown in Figure~\ref{fig:alphaCensky}.

\subsection{Photogravitational Assists in the Solar Neighborhood}
\label{sub:pga}

\begin{figure*}
\includegraphics[width=.496\linewidth]{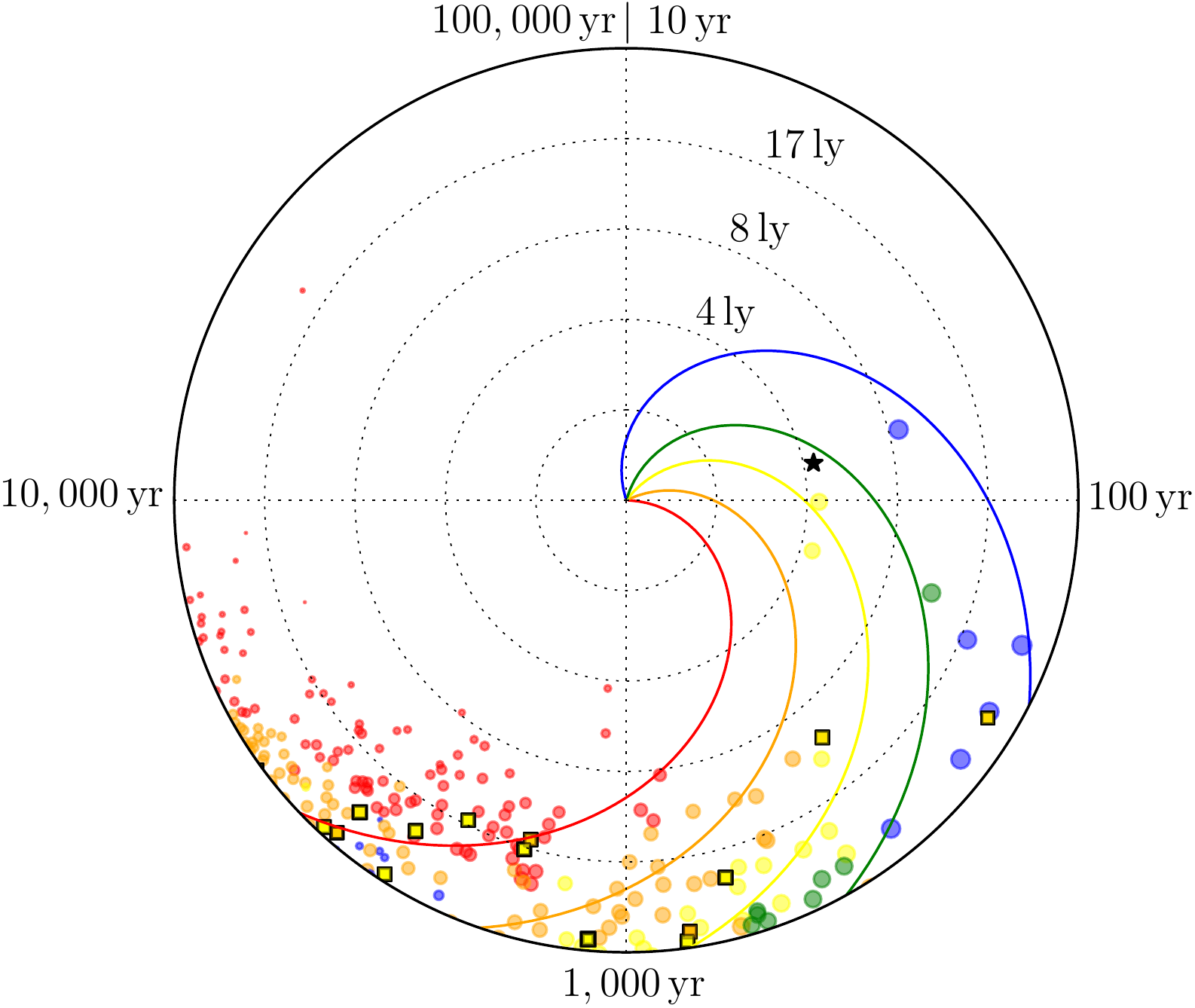}
\hspace{.12cm}
\includegraphics[width=.497\linewidth]{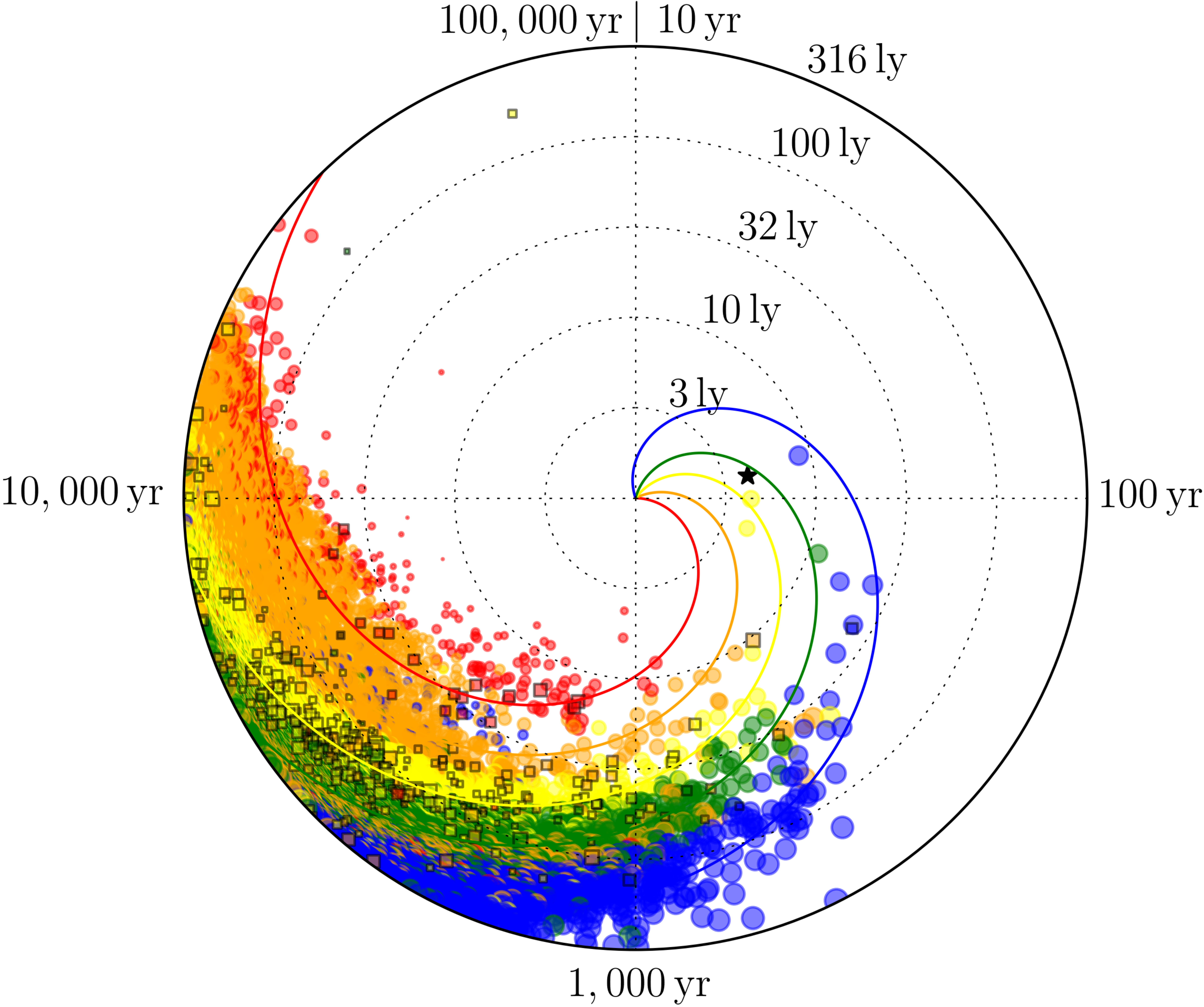}
\caption{\label{fig:clock}Travel times (hour angles) versus stellar distance to the solar system (radial coordinate) for stars in the solar neighborhood. Symbols refer to numerical trajectory simulations to individual targets; lines illustrate logarithmic spirals as per Equation~\eqref{eq:logspir} {for $r_{\rm min}=5\,R_\star$}. These spirals are parameterized using M2V (red), K5V (orange), G2V (yellow), F3V (green), and A0V (blue) template stars \citep{2013ApJS..208....9P}. Square symbols depict stars with known exoplanets. The black star symbol denotes $\alpha$\,Cen. {\it Left}: \ All 117 stars within 21\,ly around the Sun. {\it Right}: \, 22,683 stars out to 316\,ly around the Sun.}
\end{figure*}

Moving on to other stars in the solar neighborhood, Figure~\ref{fig:clock} illustrates our findings for the minimum possible travel times of a graphene-class lightsail to perform a photogravitational assist to a full stop, i.e., into a bound orbit around the respective star. The left panel shows 117 stars within 21\,ly, and the right panel shows 22,683 stars out to 316\,ly. The meanings of the symbols and lines are described in the figure caption. For these full-stop maneuvers to work, note that the injection speeds need to be achieved at departure from the solar system in the first place.

As an intriguing result of this study, we find that Sirius\,A, though being about twice as far from the Sun as $\alpha$\,Cen, could actually decelerate an incoming lightsail to a full stop after only about 69\,yr of interstellar travel ($v_{\infty, {\rm max}}=12.5\%\,c$).\footnote{\citet{2017ApJ...835L..32H} calculated a value of $v_{\infty, {\rm max}}~=~14.9\%\,c$. They used $r_{\rm min}=5\,R_\star$ without any constraints on the maximum effective temperature of the sail to prevent overheating, see Equation~\eqref{eq:heating}.} This is due to the star's particularly high luminosity of about 24.2 solar luminosities. The derived travel time compares to a minimum travel time of 75\,yr to $\alpha$\,Cen if a sequence of photogravitational assists is used at stars A and B to perform a full stop, and it compares to 101\,yr of interstellar travel to $\alpha$\,Cen\,A or 148\,yr to $\alpha$\,Cen\,B if only the target star is used for a slowdown to zero. The case of Sirius is particularly interesting because it is actually a binary system and Sirius\,B is a white dwarf \citep{1844MNRAS...6..136.,1915PASP...27..236A}. {Photogravitational assists in the Sirius\,AB system would need an exact determination of the binary orbit prior to launch \citep[for recent astrometry of the orbit, see][]{2017ApJ...840...70B}, which could then allow a sequence of flybys around an A1V main-sequence star and a white dwarf.}

Table~\ref{tab:tau} lists our results for the maximum injection speeds and minimum travel durations to the 10 most nearby stars shown in Figure~\ref{fig:clock} in order to perform a full stop via photogravitational braking {(the full list of 22,683 objects is available in the journal version of this article)}. {The results have been obtained using numerical trajectory simulations from our modified $N$-body integrator.} The objects are ordered by increasing travel time.

\floattable
\begin{deluxetable*}{rccccc}[h!t!]
\tablecaption{An Interstellar Travel Catalog to Use Photogravitational Assists for a Full Stop. \label{tab:tau}}
\tablehead{
\colhead{\#} & \colhead{Name} & \colhead{Travel time} & \colhead{Distance} & \colhead{Luminosity}  & \colhead{{Maximum injection speed}}  \\
        &           &     \colhead{(yr)}          &      \colhead{(ly)}     & \colhead{($L_\odot$)} &      \colhead{(${\rm km\,s}^{-1}$)}
}
\startdata
1. & Sirius\,A\tablenotemark{a} & 68.90 & 8.58 & 24.20 & 37,359 \\
2. & $\alpha$\,Centauri\,A\tablenotemark{b} & 101.25 & 4.36 & 1.56 & 12,919 \\
3. & $\alpha$\,Centauri\,B\tablenotemark{b} & 147.58 & 4.36 & 0.56 & 8863 \\
4. & Procyon\,A\tablenotemark{a} & 154.06 & 11.44 & 6.94 & 22,278 \\
5. & Altair & 176.67 & 16.69 & 10.70 & 28,341 \\
6. & Fomalhaut\,A\tablenotemark{c} & 221.33 & 25.13 & 16.67 & 34,062 \\
7. & Vega & 262.80 & 25.30 & 37.0 & 28,883 \\
8. & Epsilon Eridiani & 363.35 & 10.50 & 0.495 & 8669 \\
9. & Rasalhague & 364.9 &  46.2  &  25.81 & 37,977 \\
10. & Arcturus & 369.4 & 36.7  & 170  & 29,806\\
\enddata
\tablenotetext{a}{Host to a white dwarf companion.}
\tablenotetext{b}{Successive assists at $\alpha$\,Cen\,A and B could allow deceleration from much faster injection speeds, reducing travel times to 75\,yr to both stars (see Section~\ref{sub:optimized}).}
\tablenotetext{c}{Host to an exoplanet, a debris disk, and two companion stars, one of which shows an accretion disk itself.}
\tablecomments{Stars are ordered by increasing travel time from Earth. The hypothetical lightsail has a nominal mass-to-surface ratio ($\sigma_{\rm nom}$) of $8.6\times10^{-4}\,{\rm gram\,m}^{-2}$. Travel times for different $\sigma$ values scale as $\sqrt{\sigma/\sigma_{\rm nom}}$. The full list of 22,683 objects is available in the journal version of this article.}
\end{deluxetable*}

\section{Discussion}

\subsection{Launch Strategy and Aiming Accuracy}

A graphene-class sail could have a maximum ejection speed of about $11,500\,{\rm km\,s}^{-1}$ from the solar system if it was possible to bring it as close as five solar radii to the Sun and then initiate a photogravitational launch \citep{2017ApJ...835L..32H}. This is much less than the maximum injection speed of $17,050\,{\rm km\,s}^{-1}$ that can be absorbed by successive photogravitational assists at $\alpha$\,Cen\,A to C. If sunlight were to be used to push a lightsail away from the solar system, then its propulsion would need to be supported by a second energy source, e.g. a ground-based laser array, to fully exploit the potential of photogravitational deceleration upon arrival. A combination with sunlight might in fact reduce the huge energy demands of a ground-based laser system.

The aiming accuracy at departure from the solar system is key to a successful photogravitational assist at $\alpha$\,Cen. Hence, the position, proper motion, and the binary orbital motion of the $\alpha$\,Cen\,AB binary will need to be known very precisely at the time of departure. This is a rather delicate question as Gaia will not observe $\alpha$\,Cen\,AB at all.

The angular diameter of $\alpha$~Cen\,A is about $0.008''=8$\,mas as seen from Earth \citep{2003A&A...404.1087K,2017A&A...597A.137K}. In order for an interstellar projectile to successfully hit $\alpha$\,Cen\,A, a pointing accuracy of $<R_\star$ is key. A fiducial accuracy of $0.2\,R_\star$ translates into a funnel $<1.6$\,mas as seen from Earth. The current uncertainty of 3.9\,mas\,yr$^{-1}$ in the proper motion (pm) vector of $\alpha$\,Cen {\citep[$\mu = 3685.8 \pm 3.9$\,mas/yr;][]{2016A&A...594A.107K}} will result in an offset of 78\,mas (0.1\,AU at $\alpha$\,Cen) after a nominal 20\,yr journey, as proposed by the Breakthrough Starshot Initiative. Hence, the current knowledge of the celestial position and motion of $\alpha$\,Cen\,AB prevents an aimed orbital injection and swing-by to Proxima. That said, pm accuracies $<2$\,mas/20\,yr = 100\,$\mu$as\,yr$^{-1}$ can in principle be reached for $\alpha$\,Cen using dedicated astrometry. These observations will be key to a successful direction of an interstellar ballistic probe from Earth to $\alpha$\,Cen.

{The aiming accuracy will also be affected by the presence of interstellar magnetic fields if the sail has an electric charge. In fact, it might be impossible for a sail to prevent getting electrically charged due to the continuous collisions with the interstellar medium. Studies of the effects of magnetic effects on the interstellar trajectories of lightsails are beyond the scope of this study but they might be crucial to assess whether aiming accuracies of the order of one stellar radius at the target are actually possible.}

\subsection{Deflection during Stellar Encounter}

The limiting factor to the full leverage of the additive nature of the photogravitational effect in the $\alpha$\,Cen\,AB system is in their orbital inclination with respect to the Earth's line of sight. The optimal deflection angle to achieve maximum injection speeds at $\alpha$\,Cen\,A is $19^\circ$ (Figure~\ref{fig:alphaCen}). A and B will never come closer than $10.7^\circ$ from our point of view. If it were possible to let the incoming lightsail tack from an angle ($\lambda$), or ``from the side'', then this might allow faster departures than permitted for straight trajectories if the sky-projected AB angular separation upon arrival is $>~19^\circ$ (see Figure~\ref{fig:alphaCen}, top right). In fact, due to the binary's sky-projected proper motion of $23.4\,{\rm km\,s}^{-1}$ and given $\alpha$\,Cen\,A's barycentric tangential velocity of ${\approx}~8\pm5\,{\rm km\,s}^{-1}$ \citep{2016A&A...594A.107K} upon encounter, the incoming sail will have a minimal (${\approx}~0.1^\circ$) tangential velocity with respect to the line of sight from Earth.

We have looked at different possibilities to add an additional tangential speed component ($v_x$) to the sail and found that $\lambda=\arctan(v_x/v_\infty)$. One option that seems physically feasible, though technically challenging, would be to send the sail with a slight offset to $\alpha$\,Cen\,A, and then fire the onboard communication laser perpendicular to the trajectory for a time $t_{\rm l}$. This maneuver would result in a curved sail trajectory. Assuming that the laser energy output ($E_{\rm l}=P_{\rm l}t_{\rm l}$; $P_{\rm l}$ being the laser power) would be transformed into kinetic energy of the sail ($E_{\rm kin}$), we have

\begin{equation}
\lambda = \arctan(v_x/v_\infty) = \arctan\left( \sqrt{\frac{2 P_{\rm l}t_{\rm l}}{M}} v_\infty^{-1} \right) \ .
\end{equation}

\noindent
To deflect our fiducial sail with $v_\infty~=~17,050\,{\rm km\,s}^{-1}$ by $\lambda=1^\circ$, a 100\,W (or a $10$\,kW) laser would have to fire for a whole year (or 10 days), yielding $E_{\rm l}=8.6{\times}10^9$\,GJ. If an adequate miniature propulsion system could be implemented to change the incoming trajectory by $1^\circ$, then the gain in $v_{\rm \infty, max}$ would be up to several $100\,{\rm km\,s}^{-1}$ and the travel time from Earth to $\alpha$\,Cen\,A could be reduced by a few years.

Assuming that this maneuver shall not add more than about one gram to the total weight, we find that an energy density of $8.6{\times}10^9$\,GJ/gram is several orders of magnitudes higher than that of conventional chemical reactants or of modern lithium batteries. Only nuclear fission could possibly yield high enough energy densities, but this technology would likely add up to much more weight to feed the laser. In turn, an increased weight will reduce the tangential velocity that can be achieved through the conversion of laser power into kinetic energy. We conclude that current means of energy storage and conversion do not permit higher incoming sail speeds $v_{\rm \infty, max}$ by steering the sail onto a significantly curved trajectory.

\subsection{Sirius Afterburner}

Using numerical simulations, we investigated scenarios in which a graphene-class lightsail approaches Sirius\,A from Earth but minimizing the deceleration during approach while maximizing the acceleration after passage of $r_{\rm min}$ (set to $5.64\,R_\star$ to prevent fatal damage). We refer to this setup as the ``Sirius afterburner'' since the star is used to accelerate the space probe to even faster interstellar velocities than might be achievable with Earth-based technology and/or using solar photons.

We find that such a flyby at Sirius\,A can increase the velocity of a graphene-class lightsail by up to $27,000\,{\rm km\,s}^{-1}$ ($9.0\%\,c$) in the non-relativistic regime with deflection angles $\delta~\lesssim~20^{\circ}$. Consequently, stars at distances $d_\star~>~d_{\rm S}$  to the Sun ($d_{\rm S}$ being the Sun--Sirius distance) and within a sky-projected angle

\begin{equation}
\varphi~=~\delta{\Bigg (}1~-~\arcsin \left( \frac{d_{\rm S}}{d_\star} \right) {\Bigg )}
\end{equation}

\noindent
around Sirius (as seen from Earth) can be reached significantly faster using a photogravitational assist at Sirius. The maximum velocity boost of $9.0\%\,c$ is smaller than $v_{\infty, {\rm max}}=12.5\%\,c$, which we determined as the maximum loss of speed upon arrival at Sirius\,A to a full stop in Section~\ref{sub:pga}, as the stellar photon pressure is not acting antiparallel to the instantaneous velocity vector during flyby.

The same principle applies to other combinations of nearby and background stars, but Sirius\,A with its huge luminosity and relative proximity to the Sun is the most natural choice for an interstellar photogravitational hub for humanity.

\subsection{Particular Objects to Visit in the Solar Neighborhood}

Beyond the many single-target stars, there are other interesting objects in the solar neighborhood, which an ultralight photon sail could approach into a bound orbit after deceleration at the host star, such as

\begin{enumerate}
\item The nearby exoplanet Proxima\,b \citep{2016Natur.536..437A};
\item A total of 328 known exoplanet host stars within 316\,ly (right panel Figure~\ref{fig:clock});
\item The young Fomalhaut triple system with its enigmatic exoplanet Fomalhaut\,b \citep{2008Sci...322.1345K} and protoplanetary debris disks around Fomalhaut\,A and C \citep{1998Natur.392..788H,2014MNRAS.438L..96K};
\item The two white dwarfs Sirius\,B, located at 8.6\,ly from the Sun, and Procyon\,B at a distance of 11.44\,ly;
\item 36\,Opiuchi, consisting of three K stars and located at 19.5\,ly from the Sun, is the most nearby stellar triple;
\item TV\,Crateris, a quadruple system of T Tauri stars and located at 150\,ly from the Sun; and
\item PSR\,J0108-1431, between about 280\,ly and 424\,ly away, is the nearest neutron star \citep{1994ApJ...428L..53T}.
\end{enumerate}

Fomalhaut\,A is a moderately fast rotator with rotational speeds of about $100\,{\rm km\,s}^{-1}$ at the equator. Altair and Vega are very fast rotating stars \citep{2006ApJ...645..664A} with strongly anisotropic radiation fields. This would certainly affect the steering of the sail. An interstellar probe from Earth would approach Vega from a polar perspective. With the poles being much hotter and thus more luminous than the rest of the star, this could be beneficial for an efficient braking.

Beyond that, it could be possible to visit stars of almost any spectral type from red dwarf stars to giant early-type stars, which would allow studies of stellar physics on a fundamentally new level of detail. In principle, an adaptation of the Breakthrough Starshot concept that is capable of flying photogravitational assists could visit these objects and conduct observations from a nearby orbit. \, That said, photogravitational assists into bound orbits around single low-luminous M dwarfs imply very long travel times even in the solar neighborhood. The case of Proxima and its habitable zone exoplanet Proxima\,b is an exceptional case since this red dwarf is a companion of two Sun-like stars, both of which can be used as photon bumpers to allow a fast and relatively short travel to Proxima.

\subsection{Prospects of Building a Highly Reflective Graphene Sail}

Since the advent of modern graphene studies in the early 2000s \citep{Novoselov2004}\footnote{Awarded with ``The Nobel Prize in Physics 2010''. Nobelprize.org. Nobel Media AB 2014. Web. 5 April 2017. \url{http://www.nobelprize.org/nobel_prizes/physics/laureates/2010}}, huge progress has been made in the characterization of this material and in its high-quality, wafer-scale production \citep{Lee2014}. For example, \citet{SanchezValencia2014} presented a method to synthesize single-walled carbon nanotubes, which have interesting electronic properties that make it a candidate material for extremely light wires in the onboard electronics of a graphene-class sail. Carbon nanotubes might also be the natural choice for a material to build a rigid sail skeleton of. \citet{sorensen2016process} patented a high-yield method for the gram-scale production of pristine graphene nanosheets through a controlled, catalyst-free detonation of C$_2$H$_2$ in the presence of O$_2$.

Private companies offer cm$^2$-sized mono-atomic layers of graphene sheets at a price of about 75\,EUR, which translates into a price of 750 million EUR for a $10^5\,{\rm m}^2$ sail. If the price decline for graphene production continues its trend of three orders of magnitude per decade over the next 10 years, this could result in costs of 750,000\,EUR for the production of graphene required for a $10^5\,{\rm m}^2$ sail in the late 2020s. The availability of affordable, high-quality graphene for large structures is key to the mission concept assumed in this study. It might be necessary to send several probes for the purpose of redundancy to ensure that at least one sail out of a fleet will survive decades of interstellar travel and successfully perform the close stellar encounters.

{One key challenge that will affect the sail's performance of a photogravity assist is in the emergence of inhomogeneities of the reflectivity  across the sail area. A sail reflectivity of 99.99\% has been assumed in our calculations, but the bombardment of the interstellar medium will create tiny holes in the sail \citep{2017ApJ...837....5H} that would cause local decreases of reflectivity. Certainly, the sail would need to be able to autonomously compensate for the resulting torques during the deceleration phase, e.g. via proper orientation with respect to the approaching star.}

\section{Conclusion}

This report describes a new means of using stellar photons, e.g. in the $\alpha$\,Cen system, to decelerate and deflect an incoming ultralight photon sail from Earth. This improved method of using photogravitational assists is different from gravitational slingshots as they have been performed many times in the solar system and different from the photogravitational assists described by \citet{2017ApJ...835L..32H}, in the sense that the lightsail is not flung around the star but it rather passes in front of it. In other words, we propose that the star is not being used as a catapult but rather as a bumper. If the mass-to-surface ratio of the lightsail is sufficiently small, then photogravitational assists may absorb enough kinetic energy to park it in a bound circumstellar orbit or even transfer it to other stellar or planetary members in the system.

Proxima\,b, the closest extrasolar planet to us, is a natural prime target for such an interstellar lightsail. Its M dwarf host star has a very low luminosity that could only absorb small amounts of kinetic energy from an incoming lightsail. Its two companion stars $\alpha$\,Cen\,A and B, however, have roughly Sun-like luminosities, which means that successive photogravitational assists at $\alpha$\,Cen\,A, B, and Proxima could be sufficiently effective to bring an ultralight photon sail to rest.

In this paper, we investigate the case of a graphene-class sail with a nominal mass-to-surface ratio of $\sigma_{\rm nom}=8.6{\times}10^{-4}\,{\rm gram\,m}^{-2}$ and find that our improved ``bumper'' technique of using photogravitational assists allows maximum injections speeds of up to $v_{\infty, {\rm max}}=17,050\,{\rm km\,s}^{-1}$ ($5.7\%\,c$) at $\alpha$\,Cen\,A, implying travel times as short as 75\,yr from Earth. The maximum injection speed at Proxima is $1270\,{\rm km\,s}^{-1}$, which the sail would be left with after assists at $\alpha$\,Cen\,A and B. This residual speed means another 46\,yr of travel between the AB binary and Proxima, or a total of 121\,yr from Earth. Travel times for lightsails with larger mass-to-surface ratios ($\sigma$) scale as $\sqrt{\sigma/\sigma_{\rm nom}}$.

The exact value of $v_{\infty, {\rm max}}$ at the $\alpha$\,Cen system depends on the deflection angle ($\delta$) required by the sail to go from A to B and, hence, $v_{\infty, {\rm max}}$ depends on the instantaneous orbital alignment of the AB binary upon arrival of the sail. We performed numerical simulations of sail trajectories under the effects of both the gravitational forces between the sail and the star and the stellar photon pressure acting upon the sail to parameterize $v_{\infty, {\rm max}}(\delta)$. We then used calculations of the orbital motions of the AB stars to first obtain $\delta(t)$ and then $v_{\infty, {\rm max}}(t)$ for the next 300\,yr. This provides us with the expected travel times $\tau(t)$ and with the times of arrival at $\alpha$\,Cen\,A for launches within the next 300\,yr. In general, we find that $75\,{\rm yr}\leq\tau(t)\leq95\,{\rm yr}$ for a graphene-class sail cruising with $v_{\infty, {\rm max}}$.

A minimum in the travel time, which might be interesting for real mission planning, occurs for a departure on 8 September 2092 with a photogravitational assist at $\alpha$\,Cen\,A in late 2167 after 75\,yr of interstellar travel. The difference between the maximum and minimum travel times to permit photogravitational assists from $\alpha$\,Cen\,A via B to Proxima is only about 20\,yr, so that an earlier departure (e.g. in 2040) might entail somewhat longer travel times (e.g. 91\,yr) but still allow a much earlier arrival (e.g. 2131) than the departure near the next minimum of $\tau(t)$.

Beyond that, photogravitational assists may allow injections into bound orbits around other nearby stars within relatively short travel times. We identified Sirius\,A as the star that permits the shortest travel times for a lightsail using stellar photons to decelerate into a bound orbit. At a distance of about 8.6\,ly, it is almost twice as distant as the $\alpha$\,Cen system, but its huge power output of about 24 solar luminosities allows injection speeds of up to $v_{\infty, {\rm max}}~=~37,300\,{\rm km\,s}^{-1}$ ($12.5\%\,c$). These speeds cannot be obtained from the solar photons alone upon departure from the solar system, and so additional technologies (e.g. a ground-based laser array) will need to be used to accelerate the sail to the maximum injection speed at Sirius\,A. Beyond the compelling opportunity of sending an interstellar spacecraft into a bound orbit around Sirius\,A within a human lifetime, its white dwarf companion Sirius\,B could be visited as well using a photogravitational assist at Sirius\,A. We identify other interesting targets in the solar neighborhood that allow photogravitational assists into bound orbits, the first 10 of which imply travel times between about 75\,yr ($\alpha$\,Cen\,AB) and 360\,yr (Epsilon Eridiani) with a nominal graphene-class lightsail.

Many of the technological components of the lightsail envisioned in this study are already available, e.g. conventional spacecraft to bring the lightsails into near-Earth orbits for sun- or laser-assisted departure. Other components are currently being developed, e.g. procedures for the large-scale production of graphene sheets, nanowires with the necessary electronic properties consisting of single carbon atom layers, gram-scale cameras and lasers (for communication between the sail and Earth), or sub-gram-scale computer chips required to perform onboard processing etc. We thus expect that a concerted effort of electronic, nano-scale, and space industries and research consortia could permit the construction and launch of ultralight photon sails capable of interstellar travels and photogravitational assists, e.g. to Proxima\,b, within the next few decades.

\vspace{-0.2cm}

\acknowledgments
{The authors thank an anonymous referee for her or his valuable report.} This work was supported in part by the German space agency (Deutsches Zentrum f\"ur Luft- und Raumfahrt) under PLATO Data Center grant 50OO1501 {and it} made use of NASA's Astrophysics Data System Bibliographic Services. {This work has made use of data from the European Space Agency (ESA)
mission {\it Gaia} (\url{https://www.cosmos.esa.int/gaia}), processed by the {\it Gaia} Data Processing and Analysis Consortium (DPAC;
\url{https://www.cosmos.esa.int/web/gaia/dpac/consortium}).}

\bibliographystyle{yahapj}
\bibliography{references}

\end{document}